\documentclass[11pt]{article}
\usepackage{article,epsfig}

\def\Journal#1#2#3#4{{#1} {\bf #2}, #3 (#4)}

\def\APJ{\em ApJ}

\def\AA{\em A\&A}
\def\SSR{\em Space Sci. Rev.}

\def\be{\begin{equation}}
\def\ee{\end{equation}}
\def\bea{\begin{eqnarray}}
\def\eea{\end{eqnarray}}

\begin{document}
\vspace*{4cm}
\title{STATISTICAL INVESTIGATION OF THE NON-THERMAL EMISSION OF GALAXY CLUSTERS}

\author{J. LANOUX, E. POINTECOUTEAU, M. GIARD, L. MONTIER}

\address{Centre d'Etude Spatiale des Rayonnements, CNRS/Universit\'e de Toulouse, 9 Avenue du colonel Roche, BP 44346, 31028 Toulouse Cedex 04, France}

\maketitle\abstracts{
A diffuse non-thermal component has now been observed in massive merging clusters. To better characterise this component, and to extend analyses done for massive clusters down to a lower mass regime, we are conducting a statistical analysis over a large number of X-ray clusters (from ROSAT based catalogues). By means of their stacked radio and X-ray emissions, we are investigating correlations between the non-thermal and the thermal baryonic components. We will present preliminary results on radio-X scaling relations with which we aim to probe the mechanisms that power diffuse radio emission ; to better constrain whether the non-thermal cluster properties are compatible with a hierarchical framework of structure formation ; and to quantify the non-thermal pressure.}

\section{Introduction}

Galaxy clusters are the largest gravitationally bound systems of the Universe. Most of their mass is in the form of dark matter ($\sim 85\%$). A few percents is in the form of galaxies. The rest ($\sim 12\%$) is in the form of diffuse hot gas, which is the intracluster medium. This gas cools down via thermal Bremsstrahlung emission observed at X-ray wavelengths. But diffuse radio sources were also observed in $\sim50$ massive and merging clusters. Their radio emission is related to highly relativistic electrons and magnetic fields that populate the intracluster space. These non-thermal components are observed at radio wavelengths due to their synchrotron emission. To date, there are two main hypothesis to explain their origin: violent acceleration processes such as mergers occurring during clusters lifetime, and secondary electrons injected during proton-proton collisions. We observe three different types of diffuse radio sources in galaxy clusters: halos, relics and mini-halos (e.g. Ferrari et al. 2008~\cite{ferrari} for a review).
The non-thermal component contributes to the total pressure of the intracluster medium and can bias mass proxies up to $\sim 15\%$ (Dolag et al. 2000~\cite{dolag}).

Gas properties are following dark matter properties (e.g. Arnaud 2005~\cite{arnaud} for a review). Studying the correlation between the thermal and the non-thermal components then allow us to observe whether the population of electrons is following the gas, and so the dark matter in the process of structure formation. Liang et al. (2000)~\cite{liang} were the first to show a correlation between the radio and X-ray luminosities in galaxy clusters. This correlation was confirmed by subsequent works such as Giovannini et al. (2009)~\cite{gio} or Brunetti et al. (2009, B09 hereafter)~\cite{bru}. But this correlation has only been highlighted for samples of massive clusters with radio halos or relics and showing features of merging processes.

In this work, we study this correlation down to a lower mass regime of clusters. So far, we have been limited by the sensitivity of radio observations. For instance, Cassano et al. (2008)~\cite{cass} searched for radio halos in clusters using the NVSS and the GMRT. They found that for clusters with X-ray luminosities higher than $10^{45}$ erg/s, only 40\% showed radio halos, and that this percentage decreases rapidly with lower X-ray luminosities. Moreover, radio halos and relics have only been found in merging clusters, i.e. in clusters dynamically disturbed. However, it is not trivial to derive the dynamical state of clusters from available data. Therefore, we investigated the correlation via a statistical approach. We applied statistical and stacking tools to a X-ray (RASS) and a radio (NVSS) all-sky surveys, working with large cluster catalogues derived from ROSAT.

We used the following $\Lambda$CDM cosmology: $H_0 = 70\,km\,s^{-1}\,Mpc^{-1}$, $\Omega_m = 0.3$ and $\Omega_{\Lambda} = 0.7$.
 
\section{Data sets and catalogues}

We made use of a X-ray and a radio survey : the ROSAT All Sky Survey (RASS) in the 0.1-2.4 KeV band and the NRAO VLA Sky Survey (NVSS) at 1.4 Ghz that covers the entire sky north of a declination of $-40^{\circ}$. 

We used a compilation of X-ray cluster catalogues assembled by Piffaretti et al. (2010)~\cite{pi}. This compilation was homogenised to give physical quantities within an over-density $\delta_{500}$ such as $R_{500}$, $L_{500}$, $M_{500}$, etc. This compilation is based on RASS catalogues such as NORAS, REFLEX, NEP, BCS,... and serendipitous catalogues from ROSAT PSPC such as 400SD, SHARC,... We used 1,428 clusters of this compilation for which we produce estimates of $L_{radio}$ and $L_X$.

\section{Methodology}

We derive $L_{X}$ and $L_{radio}$ from the RASS and NVSS using the following steps :

\begin{itemize}
\item In order to weight properly the luminosities, we masked the brightest point sources outside an aperture of $R_{500}$ around the location of each cluster of our list  using associated catalogues: the Bright and Faint Source Catalogues for the RASS, and the NVSS Source Catalogue for the NVSS.
\item We then computed count rates and radio fluxes within $R_{500}$ taking into account the local background. For the radio fluxes, we derive as well an estimate of the radio luminosities after the removal of point sources within $R_{500}$.
\item We derived X-ray and radio luminosities within $R_{500}$. Assuming a standard self-similar evolution (Arnaud 2005~\cite{arnaud}), we scaled all luminosities  as $L_X \propto n_e^2V \propto h^{-1}(z)$ and $L_{radio} \propto n_eV \propto h(z)$ with $h(z) = h_0 \sqrt{(1+z)^3 \Omega_M + \Omega_{\Lambda}}$.
\item From the computed X-ray and radio luminosities, we draw the $L_{radio} - L_X$ correlation using bisector methods taking into account the errors in both directions.
\end{itemize}

\section{Sanity Checks}

\begin{figure}
\begin{center}
\psfig{figure=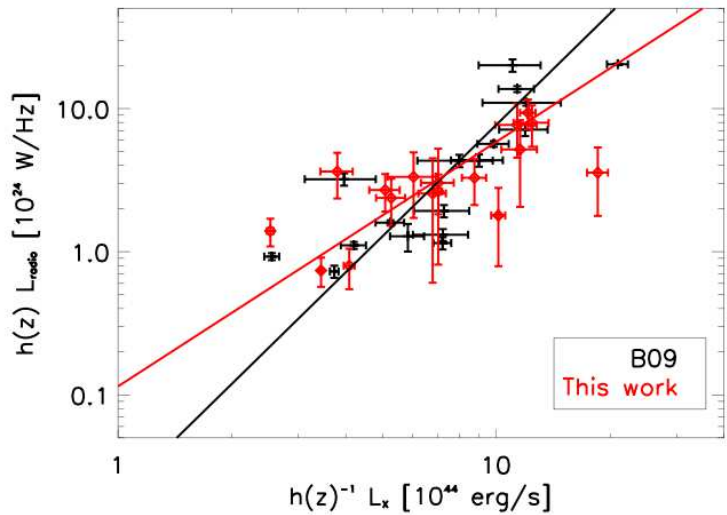,height=3in}
\end{center}
\caption{Distribution of  20 clusters from the sample of B09 in the radio-X luminosity plane. Black points mark luminosities from their work, red points mark our computed luminosities for the same clusters. Solid lines show the corresponding $L_{radio} - L_X$ best fits.
\label{fig:fig1}}
\end{figure}

In order to double-check the validity of our X-ray and radio luminosities, we performed a series of sanity checks and tests. We compared our X-ray luminosities with those of Piffaretti et al. (2010)~\cite{pi}. Beside a few outliers, we are in very good agreement with their luminosities.
The cross-check of our radio luminosities was more problematic as we don't have large samples. We compared them with the radio luminosities of 20 out of 24 clusters from the sample of B09~\cite{bru}, and we are compatible with their luminosities.

We then reproduced the computation of our X-ray and radio luminosities for 1,500 random positions on the sky with $|b| > 15^{\circ}$. To derive those, we attributed an artificial redshift, $R_{500}$ and $kT_{500}$ accordingly to the respective distributions of those quantities in our list of clusters. We performed F-tests and Kolmogorov-Smirnov tests between the distributions of clusters and of random fields. The results of both tests are compatible with 0, and the distributions are therefore different.

Finally, we drew the $L_{radio} - L_X$ correlation for the random fields, and we found a flat correlation. Nevertheless, the fit showed a nonzero mean radio background. Thus, we subtracted it from all our luminosities (Figure~\ref{fig:fig2}, right panel).

\section{$L_{radio} - L_X$ correlation}

\begin{figure}
\begin{center}
\psfig{figure=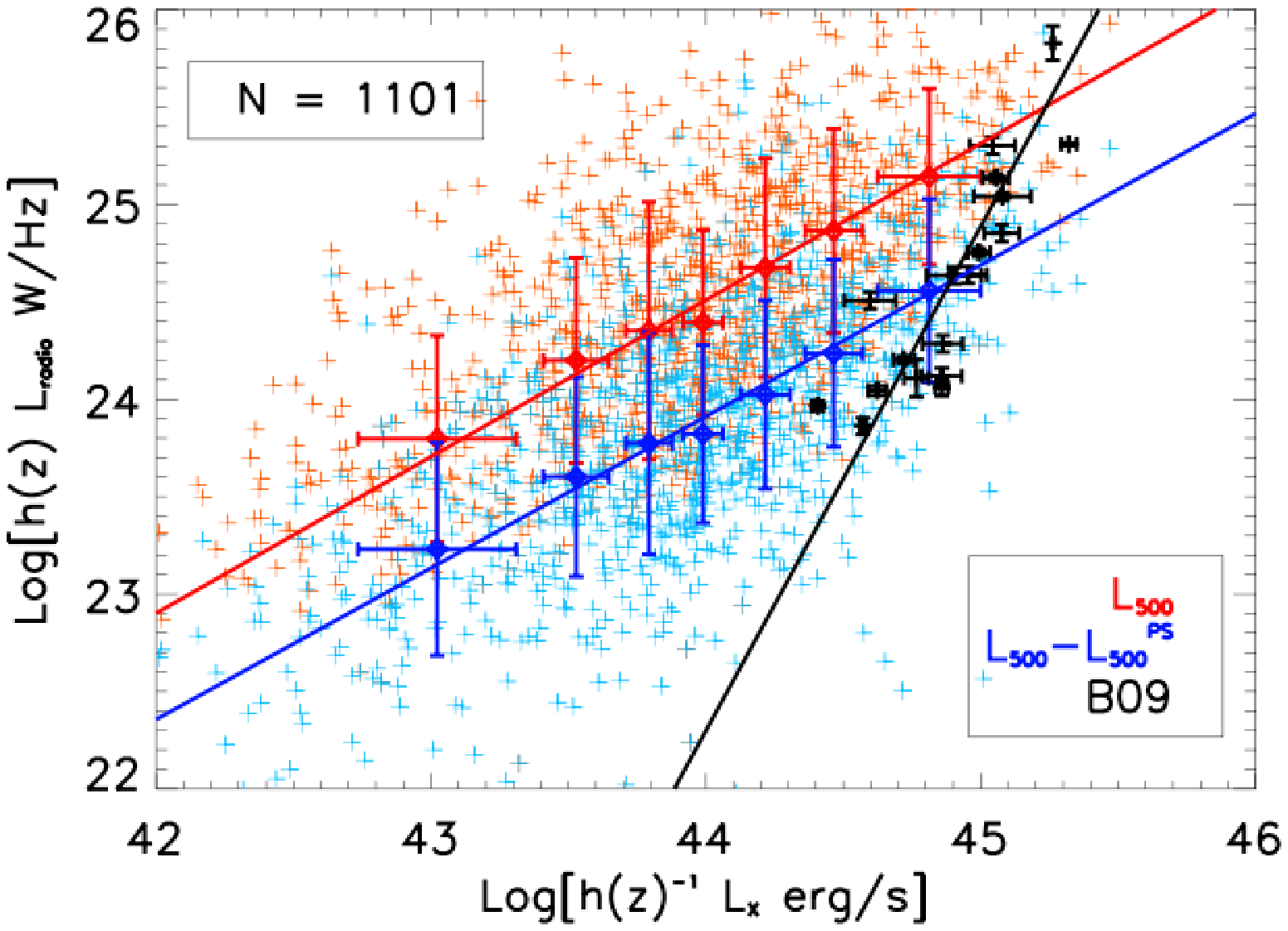,height=2.2in}
\psfig{figure=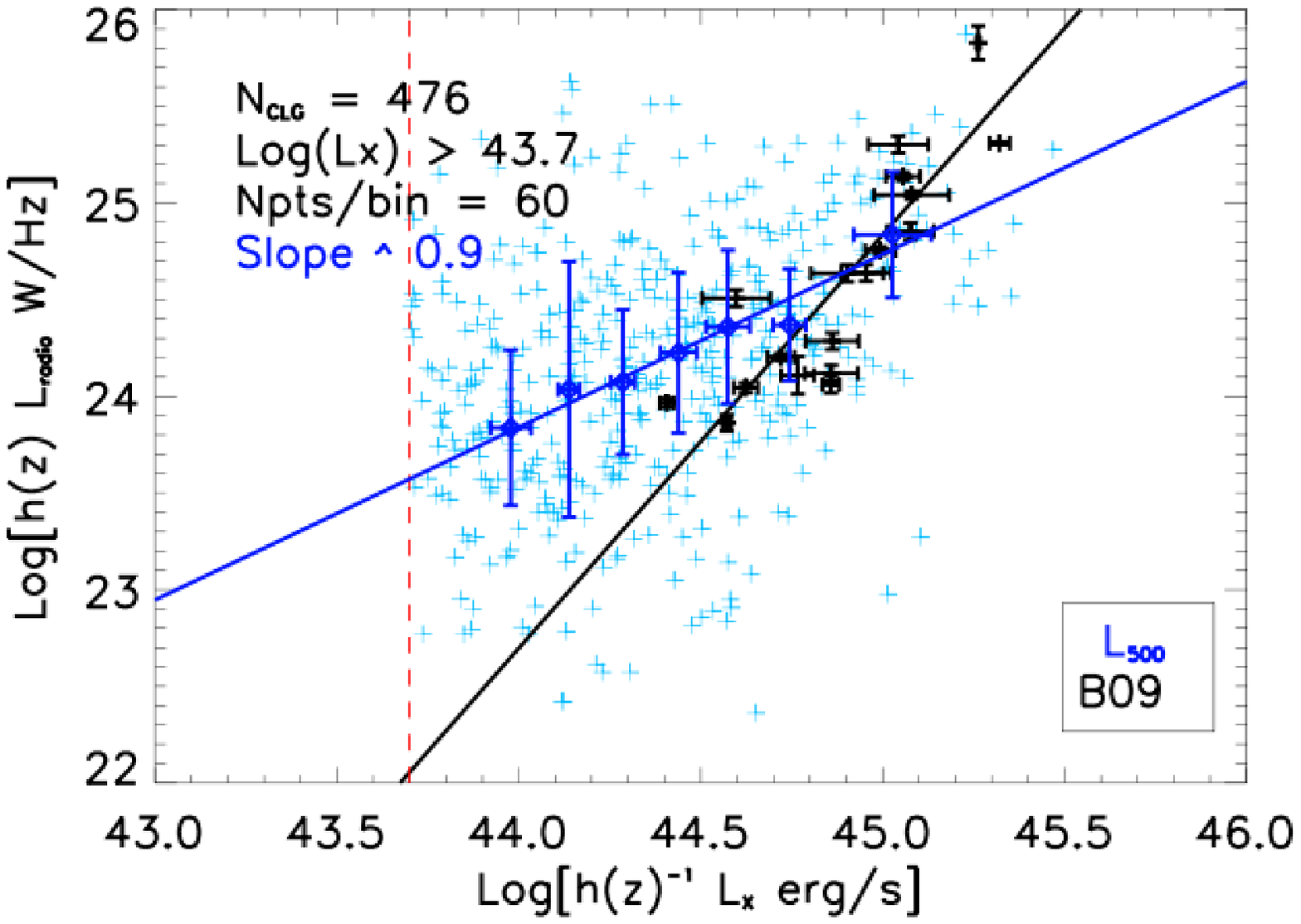,height=2.2in}
\end{center}
\caption{Distribution of clusters in the radio-X luminosity plane for clusters of our list (\textit{left panel}) and for clusters with X-ray luminosity higher than $5\times10^{43}$ erg/s (\textit{right panel}). Red crosses mark individual radio luminosities with all the fluxes within $R_{500}$, blue crosses mark individual radio luminosities with the fluxes subtracted from known point sources, and black points mark clusters from the sample of B09. Diamonds mark binned luminosities. Solid lines show the corresponding $L_{radio} - L_X$ best fits.
\label{fig:fig2}}
\end{figure}

As a first step, we reproduced the work of B09. We plotted their clusters and correlation. We then identified those clusters in our list of clusters and we plotted them as well as the corresponding correlation using our computed luminosities (Figure~\ref{fig:fig1}). We derived a slope of $\sim 1.7$. B09 derived a slope of $\sim 2.6$ for luminosities corrected from evolution ($\sim 2.2$ with no evolution). We are marginally compatible with their results. However, it matches very well the slope of $\sim 1.7$ derived with linear methods (e.g. Feretti et al. 2005~\cite{feretti} for 16 clusters, Giovannini et al. 2009~\cite{gio} for 33 clusters).

In a second step, we drew the $L_{radio} - L_X$ correlation for the clusters of our list (Figure~\ref{fig:fig2}, left panel). We compared the results obtained with the two different radio luminosities that we computed (luminosities with all the fluxes within an aperture of $R_{500}$, and luminosities subtracted from known point sources). We found the same slope, but with a lower normalization for the \textit{cleansed} luminosities. We then drew the correlation for a sub-selection of 476 clusters with $L_X > 5\times10^{43}$ erg/s ($M >~\sim 10^{14} M_{\odot}$). We then binned our X-ray luminosities in bins of equal number of clusters, i.e. 60. We fitted the binned data and derived a slope of $\sim 0.9$.

\section{Discussion and Conclusion}

We have investigated the $L_{radio} - L_X$ correlation for clusters of galaxies. We extended this analysis down to a lower mass regime ($M > 10^{14} M_{\odot}$) with respect to previous works dealing with limited samples of massive and dynamically disturbed clusters.

We found a correlation with a flatter slope ($\sim 0.9$) compared to pre-cited works. This is likely to be explained by the fact that our radio luminosities contains a component of unresolved point sources. In other terms, all the point sources (i.e. AGN) are not subtracted within $R_{500}$. This population of unresolved AGN (in the used radio data, i.e. NVSS data) may be the main contributor to our radio luminosities on the top of a diffuse emission linked to the intracluster medium.

To go further in this analysis, we need to quantify the contribution of AGN in the radio emission of galaxy clusters. This will allow us to constrain the population of AGN in dense media ; to characterize diffuse radio emissions of clusters down to the low mass regime ; and to probe the mechanisms that power them.

\section*{Acknowledgments}

We are deeply indebted to Rocco Piffaretti for providing us with his homogenised compilation of X-ray catalogues with which this work was carried out. JL, EP and LM acknowledge the support of grant ANR-06-JCJC-0141.



\section*{References}
\begin{thebibliography}{99}


\bibitem{arnaud}
Arnaud, M., \textit{Background Microwave Radiation and Intracluster Cosmology} (2005) \texttt{[arXiv:astro-ph/0508159]}

\bibitem{bru}
Brunetti, G., Cassano, R., Dolag, K., and Setti, G., \Journal{\AA}{507}{661}{2009}

\bibitem{cass}
Cassano, R., Brunetti, G., Venturi, T., et al., \Journal{\AA}{480}{687}{2008}


\bibitem{dolag}
Dolag, K., Schindler, S, \Journal{\AA}{364}{491}{2000}

\bibitem{ferrari}
Ferrari, C., Govoni, F., Schindler, et al., \Journal{\SSR}{134}{93}{2008}

\bibitem{feretti}
Feretti, L., \textit{X-Ray and Radio Connections} (2005b) \texttt{[arXiv:astro-ph/0406090]}

\bibitem{gio}
Giovannini, G., Bonafede, A., Feretti, L., et al. \Journal{\AA}{507}{1257}{2009}

\bibitem{liang}
Liang, H., Hunstead, R. W., Birkinshow, M., and Andreani, P.,  \Journal{\APJ}{544}{686}{2000}

\bibitem{pi}
Piffaretti, R., et al.,  \Journal{\AA}{}{(to be submitted)}{2010}






\end{thebibliography}
\end{document}